\documentstyle[12pt,fleqn,cite]{article}
\newcommand{\sect}[1]{\setcounter{equation}{0}\section{#1}}

\def\sxn#1{\sect{#1}}

\def\be{\begin{equation}}
\def\ee{\end{equation}}
\def\bea{\begin{eqnarray}}
\def\eea{\end{eqnarray}}
\def\pa{\partial}

\begin{document}
\begin{titlepage}
\begin{center}
\hfill hep-th/9709061\\
\hfill CERN-TH/97-194\\
\hfill LPTENS-97/29\\
\hfill MRI-PHY/p970511\\
\vskip .2in

{\Large \bf BPS States in $N=3$ Superstrings}
\vskip .5in

{\bf Costas Kounnas$^*$\footnote{On leave from Ecole Normale
Sup\'erieure, 24 rue Lhomond, F-75231, Paris Cedex 05,
France.} and Alok Kumar$^{\dagger, \>\dagger\dagger}$}\\
\vskip .1in
{\em * Theory Division, CERN,\\
CH-1211, Geneva 23, SWITZERLAND\footnote{e-mail addresses:
kounnas@nxth04.cern.ch, kumar@iop.ren.nic.in, kumar@mri.ernet.in}}
\vskip .7cm
{\em $\dagger$ Mehta Research Institute,\\
Allahabad 221 506, INDIA}
\vskip .7cm
{\em $\dagger\dagger$ Institute of Physics,\\
Bhubaneswar 751 005, INDIA}

\end{center}

\begin{center} {\bf ABSTRACT}
\end{center}
\begin{quotation}\noindent
\baselineskip 10pt


The $N = 3$ string models are special solutions of the type II
perturbative string theories. We present explicit expressions for the
helicity supertraces,
which count the number of the perturbative BPS multiplets. Assuming
the
non-perturbative duality $(S\leftrightarrow T)$ of the heterotic
string on
$T^6$ and type II on $K_3\times T^2$ valid in $N = 4$ theories, we
derive
the $N = 3$
non-perturbative BPS mass formula by ``switching off" some of the
$N = 4$
charges and ``fixing" to special values some of the $N = 4$ moduli.
This
operation corresponds to a well-defined $Z_2$ projection acting
freely
on the compactification manifold. The consistency of this projection
and the precise connection of the $N = 4$ and $N = 3$ BPS spectrum is
shown
explicitly in several type II string constructions. The heterotic
$N = 3$
and some asymmetric type II constructions turn out to be
non-perturbative with the $S$ moduli fixed at the self-dual point
$S=i$. Some of the non-perturbative $N = 3$ type II are defined in
the
context of F-theory.

The bosonic sector of the $N=3$ string effective action is also
presented. This part can be useful for the study of 4d black holes in
connection with the  asymptotic density of BPS states in $N = 3$
string theory.

\end{quotation}
\vskip .2in
CERN-TH/97-194\\
August 1997\\
\end{titlepage}
\vfill
\eject


\sxn{Introduction}

Recently    significant progress has been made in
the study of the non-perturbative aspects of string theory. One of
the
major steps in this direction has been the identification of the
solitonic states, predicted through BPS mass formulae, with the
D-brane
states in string theory\cite{polch}. The counting of these states,
through an open string construction with Dirichlet boundary
condition,
has been shown to give expressions for  the black hole entropy,
in a number of
cases from the microscopic point of view, which match  the
macroscopic description. As a result, apart from  having possible
interesting phenomenological applications, these developments also
have
a potential to solve the Hawking paradox for the black-hole
information
loss.

A major requirement in the study of the non-perturbative aspects
of string theory is the preservation of a part of the original
supersymmetry. This, in many cases, protects the physical
quantities from receiving
quantum corrections, through non-renormalization theorems and a
semi-classical analysis is sufficient for getting exact results.

However, the major thrust of these studies has been restricted
to the case of $N=4$ \cite{senijm,kiritsis} and $N=2$ \cite{kachru}
space-time supersymmetries. The $N=4$ string theories provide the
simplest non-trivial possibilities, since the only allowed massless
matter
multiplets in this case are the vectors. The perturbative moduli
space
is parametrized by a coset, $SO(6, 22)/SO(6)\times SO(22)$. This
coset
gives a complete classification of the perturbative string spectrum
with $N=4$ supersymmetry. There are, in addition, the axion-dilaton
moduli as well, which parametrize a coset space, $SU(1, 1)/U(1)$.
The
$SU(1, 1)$ symmetry mixes the electric charges with the magnetic ones
and turns the weak-coupling string theory into a strong-coupling
one.
The BPS states of the $N=4$ theory preserve either one-half or
one-quarter of the supersymmetry. For instance, the extremal black
holes
of \cite{sen94} are  examples of  BPS states preserving
one-half
of the supersymmetry.  On the other
hand, $N=2$ supersymmetric theories
have a much richer moduli structure which can among other things
account  for the confinement in the supersymmetric gauge theories.

In this paper, we discuss some aspects of a similar study for
$N=3$ superstring theories\cite{ferkoun}.
Although, these theories have
attracted much less interest, compared with the
ones discussed above, they seem to possess in some ways
interesting features of both of $N=2$ and $N=4$ theories.
Since the physical degrees of freedom in a
vector multiplet of an $N=3$ theory are the same as those
in an $N=4$ one, the moduli space of the two theories share
similar universal properties. In particular,
the number of vectors in the spectrum still determines the scalar
manifold uniquely. Due to a similarity in vector multiplets
in the two cases, the $N=3$ theories with global supersymmetries get
automatically extended to $N=4$. This is one of the reasons for
comparatively less
attention being paid to them. On the other hand, the supergravity
sector of an $N=3$ theory resembles that of the $N=2$ case, as there
are no scalars in this sector in both cases.
The full moduli space for
$N=3$ theories for $n$ matter multiplets, at a generic point,
has the form $SU(3, n)/U(1)\times SU(3) \times SU(n)$.
The supergravity couplings therefore clearly
distinguish the $N=4$ and $N=3$ cases.
The present study highlights some of these similarities and
differences. In particular, in section 2
of this paper, we study the particle spectrum of $N=3$ string
theory with an emphasis on the BPS states. We present the
$N=3$ constructions and derive the expressions for
various helicity-supertraces that count the number of
these multiplets.

In the $N=2$ models, it is known that, depending on the
details of the string construction, the dilaton
belongs either to a vector or to a hyper-multiplet\cite{vawit}.
In the $N=3$ case
also, as we will discuss, there are two allowed projections
for constructing models. The
first possibility is to carry out a $Z_2$ projection either in
the heterotic
or the $(4, 0)$ type II models, in which  the dilaton is
projected out. This is possible,
provided one is at a self-dual point on the space of string
coupling for $N=4$. Such theories therefore are only defined at a
non-perturbative level. However, we would like to emphasize that our
projection can be used to find out both the massless and the
massive BPS spectra in these cases. One can also use a $Z_2$
projection in the type II models with $(2,
2)$
supersymmery; in this case the dilaton survives the projection
and belongs to the
vector
multiplet. We show the exact connection between these two
projections,
through a known prescription for the construction of type II dual
pairs\cite{seva}. The action of the projections on the 16 extra
right-moving coordinates of the heterotic string theory is also
obtained by examining the transformation of the twisted sector states
in the original $(2, 2)$, type II, model.

Our results therefore provide an example of duality in the $N=3$
context. This is similar to the F-theory/heterotic string
duality\cite{vafa}
discussed in the literature. There are many known
examples in the F-theory side, which are only non-perturbatively
defined\cite{dasmuk},
but can still be shown to be dual to a heterotic string
construction. In those cases, as in ours, the coupling
constants are frozen to a fixed value in the orbifold limit.
Following a similar line of study,
it may also be possible to define the $N=3$ model as a
geometric compactification of a ``hidden'' theory, which would allow
us
to study their non-perturbative aspects.

After the identification of the projection in the
heterotic (or  $(4, 0)$ type II) models, the BPS mass
formula for $N=3$ string theory is obtained through a projection
of the $N=4$ formula. We show that the expressions for
the BPS formula for the $N=4$ string have a unique truncation,
which defines the $N=3$ case, irrespective of whether the BPS
states preserve the $1/2$ or the $1/4$ of the original
supersymmetry. This uniqueness is
explained by the fact that the states of the
$N=3$ strings are classified by a single central charge,
whereas in the $N=4$ case there are two such charges.
As a result, the projection acting on both these states gives
the unique short-multiplet of an $N=3$ theory. This multiplet
preserves $1/4$ of the original $N=4$ supersymmetry. The
final mass formula is $U(3, n)$-invariant
and  formally has a structure similar to the  the one
used in writing down the black-hole entropy\cite{ferrara}.

The derivation of the BPS mass formula in theories with lesser
supersymmetries, obtained by projection of a theory with a
larger number of supersymmetries, has been discussed earlier in
a different context\cite{dvv}.  In our case, these projections
can also be used to obtain the number of BPS states with
given quantum numbers, from the knowledge of the degeneracy of
such states
in the original $N=4$ theory.

Another application of our results is to write down
the $N=3$ string effective action. In this paper we present
the bosonic part of this effective action.
The effective action is also invariant under a $U(3, n)$
symmetry, which
can be used as a solution-generating technique for these
models. In particular the black-hole solutions for $N=3$
theories, carrying 14 electric charges, can be obtained
along a line similar  to the one in \cite{sen94}
for the $N=4$ case.

\sxn{$N=3$ Constructions}

\subsection{(2,1)--Type II Models}

We now start by presenting the type II $N=3$ string models
\cite{ferkoun} and write down their partition function.
All the models of
\cite{ferkoun} have been obtained by applying
projections to the $N=8$ string theories, which
preserve modular invariance as well as
the conformal symmetries on the string worldsheet.
The $N=8$ model is described in the bosonic language,
in the light-cone gauge, by 8 worldsheet
left/right-moving bosonic and fermionic coordinates,
$\psi_i^{L, R}$ and $X_i^{L, R}$ $(i = 1,...,8)$.
In our notation, the coordinates
$\psi_{\mu}^{L, R}$ and
$X_{\mu}^{L, R}$ $(\mu = 1, 2)$ represent the space-time
degrees of freedom, whereas the remaining ones correspond to
the internal degrees of freedom. This bosonic description
will be appropriate for the asymmetric orbifold construction
in section 3 where we will discuss various issues related to the
non-perturbative BPS spectrum.

In the fermionic
construction\cite{abk},\cite{klt}, $X_i^{L, R}$'s $(i=3,...,8)$
are replaced by a pair of
Majorana--Weyl spinors $\omega_a^{L, R}$ and $y_a^{L, R}$,
$(a=1,...,6)$. To follow the standard notation of
the fermionic construction\cite{ferkoun}, we also rename
the internal components of the field
$\psi_i^{L, R}$'s as $\chi_a^{L, R}$'s.
The construction of string models amounts to
a choice of boundary conditions for these fermions,
which satisfies
local and global consistency requirements. The $N=8$
model, constructed in this manner, have four
space-time supersymmetries originating from the left-moving
sector and another four from the right-moving sector.
In the language of the fermionic construction\cite{abk},
this model is constructed by
introducing three basis sets, namely $F$, which contains
all the left- and the right-moving fermions:
\be
F = [~\psi_{\mu}^L, \chi_a^L, y_a^L , \omega_a^L~|
{}~\psi_{\mu}^R, \chi_a^R, y_a^R, \omega_a^R~]
    \>\>\>(\mu =1,2; a=1,..,6),      \label{fullset}
\ee
and the basis sets $S$ and $\bar{S}$, which contain only
eight left or right-moving fermions:
\be
 S = [~\psi_{\mu}^L, \chi_a^L~] \>\>\>
\bar{S} = [~\psi_{\mu}^R, \chi_a^R~].  \label{es-es-bar}
\ee
Four of the gravitinos of the $N=8$ model belong to
sector $S$ and the other four to  $\bar{S}$. Then by applying
appropriate projections
one obtains the $N=3$ superstring model with gauge groups of various
ranks.

First, to keep the discussion simple, we avoid those models that
make a
contribution to the massless particle spectrum from the twisted
sectors
of new basis sets. As a result, all the states in these models are a
subset of those in the $N=8$ case. For  example, the first
projection
for constructing an $N=3$ model with three matter multiplets, is
specified by a
choice of fermion basis $b^H_{3}$:
\be
  b_{3}^{H} = [~\psi_{\mu}^L,
  \chi_{1, 2}^L, y_{3,...,6}^L, y_1^L, \omega_1^L ~|~
  \psi_{\mu}^R,
  \chi_{1, 2}^R, y_{3,..,6}^R, y_1^R, \omega_1^R~].
                       \label{proj1}
\ee
This applies a left--right-symmetric
$Z_2$ projection in the planes defined by the coordinates
$\chi_{3, 4}$ and $\chi_{5, 6}$.
It breaks half of the supersymmetries
in both the sides, by projecting out two of the gravitinos
from each of the sectors $S$ and $\bar{S}$. The resulting model
has $N=4$ supersymmetry with $12$ $U(1)$ gauge fields.
The local structure of the moduli space is parametrized
by a coset:
\be
{SU(1, 1)\over U(1)} \times {SO(6, 6)\over SO(6)\times SO(6)}.
\ee
In terms of orbifold construction, $b^H_3$ acts without
fixed points, making the twisted sector states heavy.

A second projection on the $N=8$ construction
yielding a $N=3$ model specified by the
basis set:
\be
  b^h_3 = [~\psi_{\mu}^L,  y_{1, 2, 3}^L,
         \omega_4^L, \chi_{5, 6}^L, y_5^L,\omega_5^L ~|~
y_5^R,\omega_5^R~].
                        \label{proj2}
\ee
This asymmetric projection, which acts
as a $Z_2$ twist on the planes
defined by $\chi_{1, 2}^L$ and $\chi_{3, 4}^L$,
breaks another one-half
supersymmetry from the left-moving sector by
projecting out one more gravitino from $S$. The resulting
model has six $U(1)$ gauge fields and the structure of
the moduli space is now  given by the coset structure:
\be
{SU(3, 3)\over U(1)\times SU(3)\times SU(3)'}.
\ee

The massless states of the model constructed above
come from various
sectors of the original $N=8$ theory. Among these,
the bosonic ones are in the sectors classified by the
 NS--NS sector $\phi$ and the R--R sector $S\bar{S}$.
These massless states can be arranged in various
representations of the subgroups of $SU(3, 3)$,
the symmetry group of the moduli deformations; the
subgroup $SU(3, 3; Z)\subset SU(3, 3)$ defines
the conjectured $\cal U$-duality group
for this $N=3$ string construction. For
convenience, we decompose the compact
subgroups $SU(3)$ and $SU(3)'$ of $SU(3, 3)$ as
\be
SU(3) \rightarrow U(1) \times SU(2)_R,
{}~~~~~~~{\rm and}~~~~~~~
SU(3)' \rightarrow U(1) \times SU(2)'_R.
\ee
The group $SU(3, 3)$  can also be decomposed as
\be
SU(3, 3) \rightarrow SU(1, 1) \times SU(2, 2).
\ee
The massless scalars from the NS--NS sector  then are:

(i) dilaton and axion, which parametrize the coset
\be
{SU(1, 1)\over U(1)}~,
\nonumber
\ee

(ii) eight scalars parametrizing
\be
{SO(2, 4)\over SO(2)_L \times SO(4)_R }~\equiv
{}~{SU(2, 2)\over SU(2)_R \times SU(2)'_R \times U(1)_L}.
\nonumber
\ee
This sector also provides two
$U(1)$ gauge fields that transform as a vector of
$SO(2)_L$. We have four additional $U(1)$ gauge fields
and four complex scalars
transforming as ${\bf (2, 1)}$ and ${\bf (1, 2)}$
of $SU(2)_R \times SU(2)'_R$ from the R--R sector.

\vskip .4cm
$N=3$ string models with a gauge sector of various other
ranks, $n$, have  been presented in \cite{ferkoun}.
For $n>3$ they involve projections whose twisted sectors
give extra contributions to the massless spectrum.
\vskip .4cm
{\bf The SU(3,3+8) model}
\vskip .4cm
An $n=11$ model constructed in \cite{ferkoun} uses the
projections
\be
  b^H_{11} = [~\psi_{\mu}^L,
  \chi_{1, 2}^L, y_{3,..,6}^L ~|~
  \psi_{\mu}^R,
  \chi_{1, 2}^R, y_{3,..,6}^R ~],
                       \label{proj1'}
\ee

\be
  b^h_{11} = [~\psi_{\mu}^L,  y_{1, 2, 3, 4}^L, \chi_{5, 6}^L,
          ~y_{5,6}^L, \omega_{5,6}^L~|~  y_{5,6}^R, \omega_{5,6}^R],
                        \label{proj2'}
\ee
and
\be
 T = [~ y_{5,6}^L, \omega_{5,6}^L~|~   y_{5,6}^R, \omega_{5,6}^R~].
\label{torus}
\ee
The basis $T$  factorizes the (5,6)-torus with independent
boundary conditions.
The  modular invariant
partition function for the above $N=3$ model is:
\be
Z^{\rm string} = {1\over {\rm Im}\tau ~\eta^{2}\bar{\eta}^{2}}~~
{1\over 4}\sum_{(H, G, h, g)}~~~
{1\over 4}\sum_{(\gamma, \delta, \gamma', \delta')}
{}~e^{i \pi (\gamma' g + \delta' h + g H)}~~ Z_L ~ Z_R ~,
\label{partition}
\ee
where $Z_{L, R}$ are themselves the products of
contributions from worldsheet fields
$\psi_{\mu}^{L, R}$, $\chi_a^{L, R}$,
$\omega_a^{L, R}$ and $y_a^{L, R}$ written in terms of
Riemann theta functions.  The boundary conditions
of these fields as
specified by the various indices in the sum above. We
then have
\be
       Z_{L} = Z_{L}^{\psi \chi}~ Z_{L}^{\omega}
                  ~Z_{L}^{y},~~~~~~~
       Z_{R} = Z_{R}^{\psi \chi} ~Z_{R}^{\omega}
                  ~Z_{R}^{y}~,
                                \label{product}
\ee
with
\be
     Z_L^{\psi \chi} =  {1\over 2}~\sum_{(a, b)}
    {(-)^{a + b + ab}\over \eta^4}
     ~~\theta[^a_b]
       ~\theta [^{a+h}_{b+g}]
       ~\theta [^{a-H-h}_{b-G-g}]
       ~\theta [^{a+H}_{b+G}]  ~,
                      \label{leftpsi}
\ee
\be
     Z_L^{\omega} = {1\over \eta^3}
     ~~\theta[^{\gamma}_{\delta}]
     ~\theta [^{\gamma}_{\delta}]
     ~\theta [^{\gamma'}_{\delta'}]  ~,
                      \label{leftomega}
\ee
\be
     Z_L^{y} =  {1\over \eta^3}
     ~~ \theta [^{\gamma + h}_ {\delta + g}]
     ~\theta [^{\gamma - H - h}_{\delta - G - g}]
     ~\theta [^{\gamma' + H}_{\delta' + G }] ~.
                      \label{leftchi}
\ee
Similarly the contributions of the right-moving fermions
is given as
\be
     Z_R^{\psi \chi} = {1\over 2}~\sum_{{\bar a},{\bar b} = 0}^1
     {(-)^{\bar{a}+\bar{b} +\bar{a}\bar{b}} \over \bar{\eta}^4}
     ~~ \bar{\theta}^2 [^{\bar{a}}_{\bar{b}}]
     ~\bar{\theta} [^{\bar{a}-H}_{\bar{b}-G}]
     ~\bar{\theta} [^{\bar{a}+H}_{\bar{b}+G}] ~,
                      \label{rightpsi}
\ee
\be
     Z_R^{\omega} = {1\over \bar{\eta}^3}
     ~~ \bar{\theta}[^{\gamma}_{\delta}]
     ~\bar{\theta} [^{\gamma}_{\delta}]
     ~\bar{\theta} [^{\gamma'}_{\delta'}]  ~,
                      \label{rightomega}
\ee
\be
     Z_R^{y} = {1\over \bar{\eta}^3}
     ~~ \bar{\theta}[^{\gamma}_{\delta }]
     ~\bar{\theta} [^{\gamma - H}_{\delta - G}]
     ~\bar{\theta} [^{\gamma' + H}_{\delta' + G }]  ~.
                      \label{rightchi}
\ee
The partition
function written in eqs. (\ref{partition})--(\ref{rightchi})
is only one of the many possibilities arising from a
choice of the GSO projection in the string construction.
One can modify these GSO projections by introducing
modular invariant phases (discrete torsions), which appear
as coefficients of  various terms in the partition function.
In our case we can choose one of the following phases
(or any product
of them):
\be
    e^{i \pi (\gamma g + \delta h + g h)}~,
{}~~~ e^{i \pi (\gamma' g + \delta' h + g h)}~,
{}~~~ e^{i \pi (\gamma G + \delta H + G H)}~,
{}~~~ e^{i \pi (\gamma' G + \delta' H + G H)}~.
\ee
The projection
$b^H_{11}$ in the partition function is represented by the twists
$H$ and $G$, while $b^h_{11}$ is represented by
$h$ and $g$. It can also
be checked that the partition function of the original $N=8$
construction is reproduced from above by setting
$H = G = h = g = 0 $ in the arguments of theta functions.
We will now use these results
to obtain an expression of the generating function
that counts the number of BPS states in
perturbative $N= 3$ string theory.
\vskip .4cm

{\subsection {Perturbative BPS States}}
\vskip .4cm

We now study the spectrum of the BPS states for
the $N=3$ model constructed above.
As mentioned before, the $N=3$ supersymmetry algebra
allows only one independent central charge. As a result
it possesses two allowed representations for the
non-vanishing value of this central charge. The long one
is a $2^6$-dimensional complex
representation of $SO(12)$ Clifford algebra and the
short one is a complex representation of dimension
$2^4$\cite{fsz},\cite{bkir}.
A vacuum configuration in a  long multiplet representation
breaks
supersymmetry completely. Since the short ones are
annihilated by one of the supersymmetry generators,
they preserve a part of the supersymmetry.
The generating functions that count the number of
$N=3$ BPS states are given
as  trace formula over the supermultiplets. In the
$N=3$ case, the helicity-generating function, defined as
\be
    Z_R (y) = {\rm str}~ y^{2 \lambda}
            = {\rm Tr} ~(-)^{2\lambda} ~y^{2\lambda},
                       \label{genrate}
\ee
with $\lambda$ denoting the helicity of the states within
a multiplet $R$, is given for a long multiplet by
\be
    Z_{\rm long} (y) = z_{[j]}~(1-y)^3 ~(1-1/y)^3.
                                   \label{long}
\ee
In eq. (\ref{long})
\be
 z_{[j]} = (-)^{2 j}
    ~~{y^{2j +1} - y^{-2j-1}\over {y - {1/y}}}~
    ~~~~ {\rm massive},
\nonumber
\ee
\be
    z_{[j]} = (-)^{2 j}~ \left(~ y^{2 j} + y^{- 2 j} ~\right)
    ~~~~~~{\rm massless},
                            \label{zedj}
\ee
for a particle of spin $j$.
For short multiplets, preserving one of the supersymmetries,
the generating function has a form:
\be
    Z_{\rm short} = 2 z_{[j]}~(1-y)^2 (1-1/y)^2.
                                        \label{short}
\ee
The extra factor of 2 in (\ref{short}) is due to the
fact that the central charge
is necessarily non-zero in this case and implies a
doubling of the representations.
The ``helicity supertrace'' over a supermultiplet $R$ is
defined as:
\be
   B_{2n}(R) \equiv {\rm str} ~\lambda^{2n} =
   {\rm Tr}_R ~[~(-)^{2\lambda} \lambda^{2n}~],
                       \label{supertrace}
\ee
and can be obtained from the generating functions $Z_R (y)$
as:
\be
   B_{2n} (R) = \left(y^2 {d\over dy^2}\right)^{2 n}~ Z_R (y)|_{y=1}.
                                     \label{derivative}
\ee
For $N=3$ supersymmetry, $B_n$  $(n < 4)$
all vanish,  $B_4$ is non-zero only for short
multiplets, and $B_6$ is non-zero for both long and short
ones. A direct computation of these quantities gives, for
$N=3$ massless multiplets:
\bea
      B_4 ({\rm vector}) &=& {3\over 2}, ~~~~~
      B_4 ({\rm sugra}) ~=~ {15\over2}, ~~~~~~~{\rm and}   \cr\cr
      B_6 ({\rm vector}) &=& {15\over 8}, ~~~~
      B_6 ({\rm sugra}) ~=~ {525\over 8}.
                           \label{b6massless}
\eea

In string theory, the above expressions are
further extended to include the infinite tower of massive states,
by defining a modified partition function \cite{bkir}:
\be
    Z^{\rm string} (v, \bar{v}) = {\rm Tr} ~q^{L_0}
{}~\bar{q}^{\bar{L_0}}
    ~e^{2\pi i v \lambda_L - 2\pi i\bar{v} \lambda_R }.
                                    \label{generate}
\ee
Such modifications to the partition function have
been studied earlier in order to obtain exact solutions of
string theory in the background of physical magnetic
fields and to investigate the associated phase-transition
phenomena\cite{kounmag},\cite{bkir}. In that context the quantities
$v$ and $\bar{v}$ play the role of the background
magnetic field. The physical helicity is given by
$\lambda = \lambda_L + \lambda_R$ and the generating
function $Z_R(y)$ for the $N=3$ supermultiplets
is obtained through an identification
$y = e^{i \pi (v + \bar{v})}$.

The helicity supertrace $B_{2n}$ in the string case can then be
derived from the generating function $Z^{\rm string} (v, \bar{v})$
by defining
\be
   Q = {1\over 2\pi i}{\partial \over \partial v},~~~~~
   \bar{Q} = - {1\over 2\pi i}{\partial \over \partial \bar{v}},
                                 \label{partial}
\ee
then
\be
    B_{2n}^{\rm string} = {\rm str} ~[\lambda^{2n} ]
           = ( Q + \bar{Q} )^{2 n}
           ~Z^{\rm string} ( v, \bar{v}) ~|_(v = \bar{v} = 0).
\ee

An explicit
expression for $Z^{\rm string} (v, \bar{v})$
for the $N=3$ string model of section 2.1 is given
by an expression that is similar to the
$v = \bar{v} = 0$ case
presented in (\ref{partition})-(\ref{rightchi}), and
has a form:
\be
   Z^{\rm string} (v, \bar{v}) = {1\over \eta^{2}\bar{\eta}^{2}}
     ~{1\over 4}\sum_{(H, G, h, g)}~~{1\over 4}\sum_{(\gamma, \delta,
     \gamma', \delta')}
    e^{i\pi (\gamma' g + \delta' h  + g H)}
               ~\xi(v)~\bar{\xi}(\bar{v})~ Z'_L ~ Z'_R,
                          \label{partition2}
\ee
where
\be
   \xi (v) = \prod_1^{\infty} { (1-q^n)^2 \over
      { (1-q^n e^{2\pi i v}) (1 - q^n e^{-2\pi i \bar{v}})}}
           = {\sin \pi v\over \pi} {\theta_1' \over \theta_1(v)},
                                  \label{xi}
\ee
is an even function of $v$:
$\xi(v) = \xi({-v})$.
The expressions for
$Z'_{L, R}$ in terms of the individual contributions of
the worldsheet fields, $\psi$, $\chi$, $\omega$ and $y$,
namely $Z'^{\psi \chi}_{L, R}$, $Z'^{\omega}_{L, R}$,
$Z'^{y}_{L, R}$,
are also identical  to the one in (\ref{product}).
However, the expressions for ${Z'}_{L, R}^{\psi \chi}$
are now modified by a change in the argument of the theta
function:
\be
     {Z'}_L^{\psi\chi} = {1\over2}~\sum_{a,b}
     {(-)^{a + b + ab} \over \eta^4}
     ~~\theta [^{a}_{ b}](v)
      ~\theta [^{a+h}_{b+g}]
      ~\theta [^{a-H-h}_{b-G-g}]
      ~\theta [^{a+H}_{b+G}]
                        \label{leftpsi2}
\ee
and
\be
 {Z'}_R^{\psi\chi} = {1\over 2}\sum_{{\bar a}, {\bar b}}
     {(-)^{\bar{a}+\bar{b} +\bar{a}\bar{b}}\over
       \bar{\eta}^4}
       ~~\bar{\theta}[^{\bar{a}}_{\bar{b}}] (\bar{v})
        ~\bar{\theta}[^{\bar{a}}_{\bar{b}}]
        ~{\bar\theta}[^{\bar{a}-H}_{\bar{b}-G}]
        ~\bar{\theta} [^{\bar{a}+H}_{\bar{b}+G}]
        ~.
                      \label{rightpsi2}
\ee

Modifications in the expression for
$Z^{\rm string} (v, \bar{v})$ in eq. (\ref{partition2}), with respect
to the one in eq. (\ref{partition}), arise from the change in the
contributions of the worldsheet fermions $\psi_{\mu}^{L, R}$
and  bosons $X_{\mu}^{L, R}$, which represent the
space-time degrees of freedom. These modifications,
due to fermions, are absorbed in
theta functions through an additional argument
$ (v )$ and $({\bar v})$. The modifications in the oscillator
contributions
from $X$ are taken into account through the extra factors
$\xi(v)$ and $\bar{\xi}(\bar{v})$.

To compute the quantities $B_{2 n}$, one can observe  that the
sum over indices $(a, b)$ and $(\bar{a}, \bar{b})$ in
eq. (\ref{partition}) involves only the
worldsheet fields $\psi_{\mu}$ and $\chi_a$ through
the terms ${Z'}_{L, R}^{\psi \chi}$ in the partition function.
We now sum over these indices and
use the Riemann identity of theta functions to write
\bea
\ {Z'}_L^{\psi \chi} &=& {1\over \eta^4}
 ~\theta[^1_1]\left({v\over 2}\right)
 ~\theta[^{1-h}_{1-g}] \left({v\over 2}\right)
 ~\theta[^{1+H+h}_{1+G+g}] \left({v\over 2}\right)
 ~\theta[^{1-H}_{1-G}] \left({v\over 2}\right) ~, \cr
             \cr
 {Z'}_R^{\psi \chi} & =&  {1\over \bar{\eta}^4}
 ~\bar{\theta}[^1_1]\left({\bar{v}\over 2}\right)
 ~\bar{\theta}[^{1}_{1}] \left({\bar{v}\over 2}\right)
 ~\bar{\theta}[^{1+H}_{1+G}] \left({{\bar v}\over 2}\right)
 ~\bar{\theta}[^{1-H}_{1-G}] \left({\bar{v}\over 2}\right) ~.
                               \label{jacobi}
\eea
To evaluate the various derivative terms for
obtaining $B_{2 n}$, we also use the properties that
among theta functions, $\theta_1$ and its
even derivatives with respect to $v$ are odd
under $v\rightarrow - v$ and
vanish at $v=0$. The odd derivatives of the
remaining theta functions, as well as that of $\xi (v)$,
are also odd under $v \rightarrow -v$ and vanish at $v=0$.
These properties simplify our calculations significantly.
For example, the RHS of
(\ref{jacobi}) immediately implies $B_2 = 0$, as expected.
The helicity supertrace $B_4$,  which counts the number of
short multiplets of $N=3$ string theory,
\be
    B_4 = ( Q + \bar{Q} )^4 ~Z (v, \bar{v})|_{v=\bar{v}=0},
                                          \label{bee4}
\ee
has a non-zero contribution only from the term
$6~ Q^2 \bar{Q}^2~ Z (v, \bar{v})|_{v=\bar{v} = 0}$
in the above expression. There
are two sectors that can give a non-zero result,
namely, the two ``$N=4$"
sectors:

(i) ${\vec h} =(h, g) = {\vec 0}$,
$~{\vec H}=(H, G) \neq \vec{0}$
and

(ii)
$ {\vec h} = {\vec H} \neq {\vec 0}$.

In the first ``$N=4$" sector
the (1,2)-complex plane remains untwisted;
the left- and the right-moving currents $J_1=\omega_1y_1 +
i\omega_2y_2$,
${\bar J}_1={\bar \omega}_1{\bar y}_1 +
i{\bar \omega}_2{\bar y}_2$
remain untwisted while the remaining ones
$J_2=\omega_3 y_3+i\omega_4 y_4$, $J_3=
\omega_5 y_5+i\omega_6 y_6$,
${\bar J}_2={\bar \omega}_3{\bar y}_3 +
i{\bar \omega}_4{\bar y}_4$,
${\bar J}_3={\bar \omega}_5{\bar y}_5 +
i{\bar \omega}_6{\bar y}_6$ are
twisted.
In the second ``$N=4$" sector the untwisted
planes are the (3,4)-left
and the (1,2)-right with untwisted currents
the $J_2$ left  and ${\bar
J}_1$ right.
The two ``$N=4$" sectors give  identical contributions.
For the first case we find:
\bea
  B_4^{{\vec h}={\vec 0}}&=&
      {3\over 4 \eta^{6} \bar{\eta}^{6}}
      ~{1\over 2}\sum_{H, G} ~~|{\theta}[^{1-H}_{1-G}]
      ~{\theta}[^{1+H}_{1+G}]~|^2  \cr
      &~&
     ~~~\times~~~{1\over 2}\sum_{(\gamma, \delta)}
     ~~|{\theta}[^{\gamma}_{\delta }]~|^4
     ~~|{\theta}[^{\gamma}_{\delta }]
     ~{\theta}[^{\gamma - H}_{\delta - G}]~|^2\cr
      &~&
     ~~~\times~~~{1\over 2}\sum_{(\gamma',\delta')}~
     |{\theta}[^{\gamma'}_{\delta'}]
     ~{\theta}[^{\gamma' + H}_{\delta' + G}]~|^2  ~.
  \label{b4final}
\eea
This expression is further simplified by using
identities involving theta functions\cite{kiritsis}.
We then get
\be
   B_4^{~{\vec h}={\vec 0}} =
 12~{1\over 2} \sum_{\gamma, \delta}
   |\theta[^{\gamma}_{\delta}]~|^4 ~
       \equiv 12 ~\Gamma_{2, 2}[^0_0]~|_{T=U=i},
                             \label{b4theta}
\ee
where in the final step we have written the
helicity trace as a lattice contribution of
signature $(2, 2)$\cite{kiritsis};
notice that both  $T$ and $U$
moduli
{\it are fixed at their self-dual points }($T=U=i$).
Adding the
contributions of the two $N=4$ sectors
(${\vec h}=(h,g)={\vec 0},~{\vec
H}=(H,G)\ne{\vec 0}$ and
${\vec h}={\vec H},~{\vec H}\ne \vec 0$), we  have
\bea
B^{\rm total}_4 &=& B_4^{~{\vec h={\vec 0},
{}~{\vec H}\ne \vec 0}} +B_4^{~{\vec h}=
{\vec H}\ne \vec 0}~\cr
 &=& 12 \Gamma_{2, 2}[^0_0]~|_{T=U=i} +
      12 \Gamma'_{2, 2}[^0_0] ~|_{T=U=i}
      ~=~24~{1\over 2} \sum_{\gamma, \delta}
          |\theta[^{\gamma}_{\delta}]~|^4.~
                         \label{b4final2}
\eea
At the massless level, eq. (\ref{b4final2})
implies $B_4({\rm massless}) = 24$, which matches
with the
combined contributions
from the supergravity sector and $11$ vector
muliplets,
written earlier in eq. (\ref{b6massless});
$B_4({\rm massless}) =
15/2+11\times 3/2=24$.

The helicity supertrace $B_6$ can also be computed in
a similar way.
In this case an analysis of the terms in the generating
function $Z^{\rm string}(v, \bar{v})$:
\be
B_6 = ( Q + \bar{Q} )^6 ~Z^{\rm string} (v, \bar{v})
                    ~|_{(v = \bar{v} = 0)},
                                 \label{bee6}
\ee
shows that
there are contributions from various sectors:

$\bullet~$ {\bf The two ``$N=4$" sectors},
$$
(i)~~~B_6^{~{\vec h}={\vec 0},~
{\vec H} \neq {\vec 0} }~~~~~
{\rm and}~~~~~
(ii)~~~B_6^{~{\vec h}={\vec H} \neq {\vec 0}}, $$
which  give a non-vanishing  contributions from
the terms
$ 15~Q^2 \bar{Q}^2~( Q^2 +\bar{Q}^2) $ in eq. (\ref{bee6}).

$\bullet$ {\bf The  ``$N=6$" sector},

$$(iii)~~~ B_6^{~{\vec h}\neq {\vec 0}, ~{\vec H}= {\vec 0}}~,
{}~~~~~~~~~~~~~~~~~~~~~~~~~~~~~~~~~~
$$
which gives a non-vanishing contribution from the term
$ 15~ Q^2 ~\bar{Q}^4$ in eq.~(\ref{bee6}).

 The extra derivative $( Q^2 +\bar{Q}^2)$ on
$B_4(v, \bar{v})\equiv Q^2
\bar{Q}^2~Z^{\rm string} (v, \bar{v})$ in  the two
$N=4$ sectors give rise
to a multiplicative factor  $\bigg(4+\chi[^H_G] + {\bar
\chi}[^H_G]\bigg)$ where $\chi[^H_G]$ are defined as:
\be
\chi[^H_G]\equiv\frac{12}{i\pi}
{}~\partial_{\tau}~{\rm log}{\theta[^{1+H}_{1+G}]\over \eta}
= {1\over 2} ~ \sum_{\gamma, \delta}~
\theta^4[^{\gamma}_{\delta}]~\left[ e^{i\pi(H+ G \gamma~)}
-e^{i\pi(G+~H \delta)}\right]~.
\ee
We will give our result for $B_6$  in terms of the
above functions
$\chi[^H_G]$
 and in terms of the ``shifted'' lattice
$\Gamma_{2,2}[^H_G]~|_{T=U=i}$:

\be
\Gamma_{2,2}[^H_G]~|_{T=U=i}~=~
{1\over 2}\sum_{\gamma, \delta}~
           |~\theta [^{\gamma +H}_{\delta +G}]~|^4~
           e^{i\pi~[\delta H + \gamma G + GH]}.
\label{2,2lat}
\ee
In terms of $\chi[^H_G]$ and $\Gamma_{2,2}[^H_G]$
the contribution of the two $N=4$ sectors is:

\be
B_6^{{~\vec h}={\vec 0},{\vec H} \neq {\vec 0} }
 ~=~ B_6^{{~\vec h}={\vec H} \neq {\vec 0}}~=~
\nonumber
\ee
\bea
 &=&{15 \over 2} \sum_{(H,G) \neq (0,0)}
\left( 1+ { \chi [^H_G] + {\bar \chi }[^H_G]\over 4} \right)
\left(~ \Gamma_{2,2}[^0_0] +
\Gamma_{2,2}[^H_G] ~\right)~|_{T=U=i}~~~~
   ~~~~\cr
&=&~30~\Gamma_{2,2}[^0_0] ~|_{T=U=i}~+~
{15 \over 2} \sum_{(H,G) \neq (0,0)}
 {\chi [^H_G] + {\bar \chi }[^H_G]~\over 4}
{}~\Gamma_{2,2}[^H_G]~|_{T=U=i}.
\eea
The final equality in the above equation follows
from the identities:
\be
\sum_{(H,G) \neq (0,0)}\chi[^H_G] =0 ,~~~~~~~~
\sum_{(H,G) \neq (0,0)} \Gamma_{2,2}[^H_G] =
\Gamma_{2,2}[^0_0]~|_{T=U=i}~.
\ee
\vskip .5cm
Finally the contribution from  the $N=6$ sector is:
\be
B_6^{~{\vec h}\neq {\vec 0}, ~{\vec H}= {\vec 0}}=
{45\over 4 }
 \sum_{(h,g)\neq (0,0)}~{{\bar \chi} [^h_g]\over
2}~\Gamma_{2,2}[^h_g]~|_{T_3~U_3}~,
			\label{b6tot}
\ee
where $T_3, U_3$ are the moduli of the third complex plane.
In the fermionic
construction  the moduli $(T_3,U_3)$ are fixed to
their self dual
points $T_3=U_3=i$. In the above expression we have
extended the
validity of the model for arbitrary $T_3, U_3$ moduli:
\bea
&&\Gamma_{2,2}[^h_g]~|_{T~U}= \sum_{m_i,n_i}
{\rm exp} \left[~{i\pi\tau
{}~|P_L(h)|^2 -i\pi\bar{\tau} ~|P_R(h)|^2~+~i\pi~g~m_1}~\right]
\cr
&&|P_L(h)|^2={|m_1~U+
(n_1 +{h\over 2})~T - m_2 + n_2~TU|^2 \over 2{\rm
Im}T~{\rm Im}U}~,~~~~~~\cr
&&|P_L(h)|^2~-~|P_R(h)|^2 =
2m_1~\left(n_1 +{h\over 2}\right)+2m_2~n_2~.
{}~~~~~~~~~~~~~~~~~~~~~
\eea
Note that  ${\bar \chi} [^h_g]$ in the
$N=6$ sector
originates from the  contribution of the $1{\rm st}$ and $2{\rm nd}$
right-moving complex planes;
these two planes are ``untwisted" on the right and
``twisted" on the
left.
The combination of the left-twisted and
right-untwisted is proportional
to ${\bar \chi} [^h_g]$. In the $N=4$ sectors
$ \chi [^h_g]$ and ${\bar
\chi} [^h_g]$
have a different origin; they appear because of
the extra derivative
operation $Q^2 +{\bar Q}^2$ on $B_4(v,{\bar v})$.

The total $B_6^{~\rm total}$ is the sum of all
contributions:
\be
B_6^{~\rm total}=~B_6^{~{\vec h}={\vec 0},
{}~{\vec H} \neq {\vec 0} }~+
{}~B_6^{~{\vec h}={\vec H} \neq {\vec 0}}~+
{}~B_6^{~{\vec h}\neq {\vec 0}, ~{\vec H}= {\vec 0}}.
                                       \label{b6total}
\ee
In the infrared limit
Im$~\tau \rightarrow\infty$ only  the massless states
give a non-zero contribution. In this limit
\bea
&& \sum_{(h,g)\neq 0}\chi[^h_g]~
\Gamma[^h_g]~|_{T,U}\rightarrow 2,
{}~~~~~~~~~ \Gamma_{2,2}[^0_0]~|_{T,U}\rightarrow 1, \cr
&& B_6^{~{\vec h}={\vec 0},~
{\vec H} \neq {\vec 0} }~\rightarrow
{75\over 2},~~~~~~B_6^{~{\vec h}=
{\vec H} \neq {\vec 0}} \rightarrow
{75\over 2},~~~~~B_6^{~{\vec h}\neq {\vec 0},
{}~{\vec H}= {\vec 0}}= {45
\over 4}.
\eea
Then $B_6^{~\rm total}({\rm Im}~\tau \rightarrow\infty)=345/4$,
which corresponds
to the contribution
of the massless fields of the  $N=3$ supergravity
together with the
contribution of  11 $N=3$ massless  vector multiplets;
$B_6({\rm massless}) =525/ 8 +11\times 15/8= 345/ 4$.
Furthermore, the contribution of the $N=6$ sector
matches (up to a factor of 2),  the
contribution of the massless fields of the $N=6$ supergravity:
\be
B_6({N=6,~ \rm sugra})=2B_6^{~{\vec h}\neq {\vec 0},
{}~{\vec H}= {\vec
0}}= {45 \over 2}.
\ee
The factor of 2 is due to the extra projection $(H,G)$
in $N=3$ theory.
In $N=6$ supergravity the massless sector is {\it uniquely}
determined by the
supergravity multiplet.

Both $B^{\rm total}_4$ and $B^{\rm total}_6$ are
then consistent with an $N=3$
supersymmetric structure with 11 vector multiplets. The
moduli space of the scalars form the K\"ahler manifold

\be
{SU(3,3+8)\over U(1)\times SU(3)\times SU(3+8)}
{\rm ~~with ~~ Kahler~ potential:~}
\nonumber
\ee
\be
K=-{\rm log} ~{\rm det}~\bigg[ i(T_{i{\bar j}}-{\bar T}_{i{\bar
j}})~-
{}~W_{i{\bar k}}{\bar W}_{{\bar k}j} \bigg].
\ee
In our case the moduli $T_{i{\bar j}}$ correspond  to the following
type II fields:

$\bullet$ the type II dilaton: $S=T_{11}$,

$\bullet$ the moduli of the $3{\rm rd}$ complex plane:
$$T_3=T_{22} ~~{\rm and} ~~ U_3=T_{33} ~,~~~~~~~~~~~~~~ ~~~~~~~~$$

$\bullet$ the Wilson lines  corresponding to the marginal
deformations
of  the currents associated to the  $(3_L,2_R)$-complex plane:

$~\bigg[(J_3)_L ~~{\rm and/or}~~ ({\bar J}_3)_L \bigg] \times
\bigg[(J_2)_R
{}~~{\rm
and/or}~~ ({\bar J}_2)_R\bigg]~~ \rightarrow ~~Y_1,~~iY_2$:
$$T_{23}=Y_1+iY_2,~~~~T_{32}=Y_1-iY_2~,$$

$\bullet$ the untwisted R--R scalars: $\rightarrow ~~
T_{12},~~T_{21},~~T_{13},~~T_{31}~,$

$\bullet$ the twisted R--R scalars: $~~~\rightarrow ~~W_{1,{\bar
k}}~, $

$\bullet$ the twisted scalars: $~~~ ~~ ~~~~\rightarrow ~~ W_{2,{\bar
k}}, ~~W_{3,{\bar k}}~.$
\vskip .3cm
The perturbative string spectrum varies with the values of  $T_3,
{}~U_3$
moduli and with the values of the two  Wilson lines $Y_1, ~Y_2$.
On the other hand the perturbative spectrum does not depends
on the values of all
other moduli. As we will see in the next chapter, using the
heterotic--type II and
type II (4,0)--type II (2,2) non-perturbative ${\cal U}$-duality map,
we find that
the heterotic dilaton is frozen at the fixed non-perturbative value
$S_H =i$. In type II
theory
$S_H$ is mapped to the frozen $T_0=i$  moduli  of the first complex
plane.
Indeed, all $N=3$ perturbative BPS states are selected by the $B_4$
helicity supertrace,
which,
as we have shown, does not display any dependence on the perturbative
moduli; it depends only on the radii of the complex planes $[(1_L,
1_R)+(2_L, 1_R)]$; the values of these  radii {\it are  fixed at
special points} in such a way that the  left-right-asymmetric
projection
defined by $\vec h$ is compatible with modular invariance.
Only in  $B_6$ is there  a non-trivial dependence on the perturbative
moduli;
it comes from the $N=6$ sector in which  the $3{\rm rd}$ complex
plane
is
untwisted.  In the  example
we gave above the moduli dependence of $B_6$ is through the
``shifted" lattice
$\Gamma_{2,2}[^h_g]~|_{T_3~U_3}$. Our result can be easily
generalized  to include  the presence of non-zero Wilson lines
$Y_1$ and
$Y_2$. In order to do that we must  perform the following
replacement  in $B^{\rm total}_6$
in eq. (\ref{b6tot}):
\be
{1\over 2} \sum_{(h,g) \neq (0,0)} {\bar \chi}[^h_g]~
\Gamma_{2,2}[^h_g]~~~~
\rightarrow ~~~ {1\over 2}\sum_{(h,g) \neq (0,0)}
\sum_{\epsilon,
{}~\xi}~ {\bar \rho}[^{\epsilon}_{\xi}]
{}~\Gamma_{2,4}[^{h,~\epsilon}_{g,~\xi}]~|_{T,U,Y_i}~,
\ee
where the shifted $\Gamma_{2,4}[^{h,~\epsilon}_{g,~\xi}]$
lattice and
the function $\rho[\epsilon, ~\xi]$ are defined as:
\be
\rho[^{h,~\epsilon}_{g, ~\xi}]~=~{1\over 2} ~
\sum_{\epsilon, ~\xi}~
\theta^2[^{1-\epsilon}_{1-\xi}]~ e^{i\pi \epsilon}~
\left[ e^{i\pi\xi~h }-e^{i\pi \epsilon~g }\right]~
\ee
\be
\Gamma_{2,4}[^{h,~\epsilon}_{g,~\xi}]=
\sum_{m_i,n_i}{\rm exp} \left[{ i\pi\tau |P_L|^2 -
i\pi {\bar \tau} ~\sum_1^4 (P^I)^2_R+
i\pi(gm_1 +\xi (Q_1-Q_2))}\right],
\ee
with
$$
|P_L|^2={|m_1U+(n_1+{h\over 2})T  - m_2 +
n_2(TU-{1\over 2}{\vec Y}{\vec Y})+(Q_1+{\epsilon \over 2})Y_1
+(Q_2-{\epsilon \over 2})Y_2|^2 \over 2{\rm Im}T~{\rm Im}U -
{\rm
Im}{\vec Y}~ {\rm Im}{\vec Y}}
$$
and
\bea
&&|P_L|^2-\sum_1^4 (P^I)^2_R -(Q_1+{\epsilon \over
2})^2-(Q_2-{\epsilon \over 2})^2 ~=~\cr
&&~=~ 2m_1(n_1 +{h\over
2})+ 2m_2n_2-Q_1^2-Q_2^2-{\epsilon^2\over 2}.~~~~~ ~~~~~~
\eea

We have therefore presented the explicit form of the helicity
supertrace formulae for an $N=3$ string construction.
Our study has been restricted to a particular $N=3$
model with 11 vector multiplets.  However similar results can
be derived for $N=3$ models with a lower number of gauge fields.
We give below  $N=3$ models with 7, 4, 3  and 1
gauge fields.

\vskip .4cm

{\bf The SU(3,3+4) model}
\vskip .4cm
In the fermionic construction one uses  the basis vectors
$F, S, {\bar S}$, and  $ b^H_{7}= b^H_{11}$, as previously
[see  eqs. (2.1) and (2.2)]. The vector
$ b^H_{7}$ implies  the symmetric  projection that is defined
by $\vec H$ [see  eqs. (2.1) and (2.2)], while the additional vector
\be
  b_7^h = [~\psi_{\mu}^L,  y_{1, 2, 3}^L, \omega_4^L
\chi_{5, 6}^L, ~y_{5,6}^L, \omega_{5,6}^L~|~  y_{5}^R, \omega_{5}^R,
y_1^R, \omega_1^R]
                        \label{proj7}
\ee
defines  the asymmetric projection denoted by $\vec h$.

As in the rank 11 model, there are  two $N=4$
sectors and one $N=6$ sector, which give
non-zero contributions to $B_4$ and $B_6$ supertraces.

For $B_4$ we have the following results:
\bea
B_4^{~{\vec h={\vec 0},
{}~{\vec H}\ne \vec 0}}&=& 12~{1\over 2} \sum_{\gamma, \delta}
          |\theta[^{\gamma}_{\delta}]~|^4=
      12 \Gamma_{2, 2}[^0_0]~|_{T=U=i}\cr
B_4^{~{\vec h}={\vec H}\ne \vec 0} &=& 6~{1\over 2}\sum_{\gamma,
\delta}~~\sum_{(H,G)\ne (0,0)} |\theta[^{\gamma}_{\delta}]~|^2
|\theta[^{\gamma+H}_{\delta+G}]~|^2  \cr
&=& 12~\Gamma_{2, 2}[^0_0]~|_{T=i,U=2i}-
6~\Gamma_{2, 2}[^0_0]~|_{T=U=i}.
\eea

At the massless level, $B_4^{\rm total}=B_4^{~{\vec h}={\vec
0}}+B_4^{~{\vec h}={\vec H}\ne \vec 0}\rightarrow 12+6=18$, which
matches the combined contributions
from the supergravity sector and $7$ vector
multiplets,
$B_4({\rm massless}) =
15/2+7\times 3/2=18$.
\vskip .4cm
For $B_6$ we find:
$$
B_6^{~{\vec h={\vec 0},
{}~{\vec H}\ne \vec 0}}
 =~30~\Gamma_{2,2}[^0_0] ~|_{T=U=i}~+~
{15 \over 2} \sum_{(H,G) \neq (0,0)}
 {\chi [^H_G] + {\bar \chi }[^H_G]~\over 4}
{}~\Gamma_{2,2}[^H_G]~|_{T=U=i},
$$
$$
B_6^{~{\vec h}={\vec H}\ne \vec 0}=
{15\over 2} \sum_{(H,G) \neq (0,0)}
{4+\chi [^H_G] + {\bar \chi }[^H_G]~\over 4}
{}~~\sum_{\gamma, \delta}~
|\theta[^{\gamma}_{\delta}]~|^2~,
{}~|\theta[^{\gamma+H}_{\delta+G}]~|^2 ~,~~~~~~~~~
$$
\be
B_6^{~{\vec h}\neq {\vec 0}, ~{\vec H}= {\vec 0}}=
{45\over 4 }
 \sum_{(h,g)\neq (0,0)}~{{\bar \chi} [^h_g]\over
2}~\Gamma_{2,2}[^h_g]~|_{T_3~U_3}~.
\ee
Some comments are in order:

$\bullet$ In the limit ${\rm Im}~\tau \rightarrow\infty$,
$B_6^{~{\vec h={\vec 0},
{}~{\vec H}\ne \vec 0}}\rightarrow 75/2$,
$B_6^{~{\vec h}={\vec H}\ne \vec 0}\rightarrow 30$ and
$B_6^{~{\vec h}\neq {\vec 0}, ~{\vec H}= {\vec 0}}\rightarrow 45/4$.
Then $B_6^{~\rm total}({\rm Im}~\tau \rightarrow\infty)=315/4$
 corresponds
to the contribution
of the massless fields of the  $N=3$ supergravity
together with the
contribution of  7 $N=3$ massless  vector multiplets;
$B_6({\rm massless}) =525/ 8 +7\times 15/8= 315/ 4$ as expected.

$\bullet$ $B_6^{~{\vec h}\neq {\vec 0}, ~{\vec H}= {\vec 0}}$
is identical to the  rank 11 model due to the universal behaviour
of the $N=6$ sector.

$\bullet$ The $T$-moduli of the first and second complex planes
is always
fixed at the self-dual point $T=i$. The $U$ moduli of the same
planes may
take several discrete values.

$\bullet$  In the  rank 7 model the $H=h=1$ twisted sector is
massive.

\vskip .4cm
{\bf The SU(3, 3) model}
\vskip .4cm
In this model there are no extra vector multiplets from the
twisted sectors. In the fermionic construction one uses the
basis vectors $F, S,  {\bar S}$ as before,  plus two extra basis
vectors   defining the symmetric projection
($\vec H$):
$$
 b^H_{3}= [~\psi_{\mu}^L,
  \chi_{1, 2}^L, y_{3,..,6}^L, ~y_{1}^L,\omega_{1}^L~|~
  \psi_{\mu}^R,
  \chi_{1, 2}^R, y_{3,..,6}^R, ~y_{1}^R,\omega_{1}^R~],
{}~~~~~~~~~~~~~~~~
                       \label{proj1-3}
$$
and the asymmetric projection $\vec h$:
\be
{}~~~~~ b^h_3=[~\psi_{\mu}^L,  y_{1, 2, 3}^L,
\omega_{4}^L \chi_{5, 6}^L,
          ~y_{5}^L, \omega_{5}^L~|~  y_{5}^R, \omega_{5}^R].
                        \label{proj23}
\ee

Here also there are two $N=4$ sectors and one $N=6$ sector.
Their contributions to $B_4$ and $B_6$ supertraces are as follows.
\vskip .4cm
$\bullet$ For $B_4$ we find:

$$
B_4^{~{\vec h={\vec 0},~{\vec H}\ne \vec 0}}=
B_4^{~{\vec h}={\vec H}\ne \vec 0}~=~
6~{1\over 2}\sum_{\gamma, \delta}~~\sum_{(H,G)\ne (0,0)}
|\theta[^{\gamma}_{\delta}]~|^2
|\theta[^{\gamma+H}_{\delta+G}]~|^2
$$
\be~~~~ ~~~~~~=12~\Gamma_{2, 2}[^0_0]~|_{T=i,U=2i}-
6~\Gamma_{2, 2}[^0_0]~|_{T=U=i}~.
\ee

At the massless level, $B_4^{\rm total}=B_4^{~{\vec h}={\vec
0}}+B_4^{~{\vec h}={\vec H}\ne \vec 0}\rightarrow 6+6=12$, which
matches
 the
combined contributions
from the supergravity sector and three vector
multiplets,
$B_4({\rm massless}) =
15/2+3\times 3/2=12$.
\vskip .4cm
$\bullet$ For $B_6$ we find:
$$
{}~~~~~ ~~~~ ~~B_6^{~{\vec h={\vec 0},
{}~{\vec H}\ne \vec 0}}
=B_6^{~{\vec h}={\vec H}\ne \vec 0}=
{15\over 2} \sum_{(H,G) \neq (0,0)}
{4+\chi [^H_G] + {\bar \chi }[^H_G]~\over 4}
{}~~\sum_{\gamma, \delta}~
|\theta[^{\gamma}_{\delta}]~|^2
{}~|\theta[^{\gamma+H}_{\delta+G}]~|^2 ~~~~~~~~~~
$$
\be
B_6^{~{\vec h}\neq {\vec 0}, ~{\vec H}= {\vec 0}}=
{45\over 4 }
 \sum_{(h,g)\neq (0,0)}~{{\bar \chi} [^h_g]\over
2}~\Gamma_{2,2}[^h_g]~|_{T_3~U_3}
\ee

At the massless level, $B_6^{\rm total}=
B_4^{~{\vec h}={\vec 0}}+B_6^{~{\vec h}={\vec H}\ne \vec 0}
+B_6^{~{\vec h}\neq {\vec 0}, ~{\vec H}= {\vec 0}}
\rightarrow 30+30+45/4=285/4$, which matches
 the
combined contributions
from the supergravity sector and three vector
multiplets,
$B_6({\rm massless}) =
525/ 8 +3\times 15/8= 285/ 4$.

Here again the contribution of the $N=6$ sector is the same as
in the previous models, thanks to the uniqueness of the $N=6$
sector. Notice also that the $T$ moduli of the first and second
complex  plane are fixed to their self-dual value.

\vskip .4cm
 {\bf The SU(3,1) model}

\vskip .4cm

This model is somewhat different from the previous ones in the
sense that it is defined by three asymmetric projections, of which
two are
acting on the left-moving gravitinos ${h_1,h_2}$ while the other
is acting on the right-moving ones ${h_3}$. All projections are
freely acting and thus there is no extra massless states coming
from the twisted sectors. The $h_1$ acts on the $2{\rm nd}$ and
$3{\rm rd}$ left moving complex planes, $h_2$ acts on the $1{\rm st}$
and $3{\rm rd}$ left-moving complex planes and $h_3$ acts on the
$2{\rm nd}$ and $3{\rm rd}$ right-moving complex planes. In the
fermionic construction one uses the basis vectors $F,S, \bar S$
as before,  as  well as the three asymmetric basis vectors,
which define the asymmetric projections $h_i$:
$$
b^{h_1} = [~\psi_{\mu}^L,~ \chi_{1, 2}^L,~ y_{ 3, 4, 5,6}^L,
          ~y_{1}^L, \omega_{1}^L~|  y_{1}^R, \omega_{1}^R],
{}~~~~~~~~~~~~~~ ~~~~~~~~~~~
                        \label{proj21}
$$
$$
b^{h_2}= [~\psi_{\mu}^L,~  y_{1, 2}^L,~ \chi_{3, 4}^L,
          ~\omega_{5,6}^L, ~y_{ 3}^L, \omega_{3}^L|~
y_{3}^R,\omega_{3}^R], ~~~~~~~~~~~  ~~~~~~~~~
                        \label{proj22}
$$
\be
{}~~~~~~~~~
b^{h_3}=[~\psi_{\mu}^R,~  \chi_{1, 2}^R,~ y_{3, 4,5,6}^R,
          ~y_{2}^R, \omega_{2}^R~|~  y_{2}^L, \omega_{2}^L].
                        \label{proj23'}
\ee
In this model there are three  $N=4$ and four $N=6$ sectors, which
give
non-zero contributions to the $B_4$ and $B_6$ supertraces.
Namely, the three $N=4$ sectors are:

\vskip .4cm
$~~~~~~~~$ 1) ${\vec h}_1={\vec 0},~
{\vec h}_2={\vec h}_3={\vec H}\ne {\vec 0} $,

$~~~~~~~~$ 2) ${\vec h}_2={\vec 0},~
{\vec h}_1={\vec h}_3={\vec H}\ne {\vec 0} $,

$~~~~~~~~$ 3) ${\vec h}_1 +{\vec h}_2={\vec 0},
{}~{\vec h}_1={\vec h}_2=
{\vec h}_3={\vec H}\ne {\vec 0} $.

\vskip .4cm

Whereas the four $N=6$ sectors are:

\vskip .4cm

$~~~~~~~~$ 1) ${\vec h}_1={\vec h}\ne {\vec 0},
{}~{\vec h}_2={\vec h}_3={\vec 0}$

$~~~~~~~~$ 2) ${\vec h}_2={\vec h}\ne {\vec 0},
{}~{\vec h}_1={\vec h}_3={\vec 0}$

$~~~~~~~~$ 3) ${\vec h}_3={\vec h}\ne {\vec 0},
{}~{\vec h}_2={\vec h}_1={\vec 0}$

$~~~~~~~~$ 4)  ${\vec h}_1={\vec h}_2={\vec h}\ne {\vec 0},
{}~{\vec h}_3={\vec 0}$

\vskip .4cm

All $N=4$ sectors give equal contributions to $B_4$:
\bea
&&B_{4}^1=B_{4}^2=B_{4}^3=3~
{1\over 2}\sum_{\gamma, \delta}~~\sum_{(H,G)\ne (0,0)}
|\theta[^{\gamma}_{\delta}]~|^2
|\theta[^{\gamma+H}_{\delta+G}]~|^2  \cr
&&~~~~ ~~~~~~=6~\Gamma_{2, 2}[^0_0]~|_{T=i,U=2i}-
3~\Gamma_{2, 2}[^0_0]~|_{T=U=i}~.
\eea

In the limit ${\rm Im}~\tau \rightarrow\infty$,
$B_4^{\rm total}=B_{4}^1+B_{4}^2+B_{4}^3\rightarrow 3\times 3=9$,
which corresponds
to the contribution
of the massless fields of the  $N=3$ supergravity
together with the
contribution of  one $N=3$ massless  vector multiplet;
$B_4({\rm massless}) =15/2 +1\times 3/2= 9$.

The $B_6$ receives contributions from the three
$N=4$ sectors as well as from the four $N=6$ sectors.
We find:
$$
B_6^{N=4,(1)}=B_6^{N=4,(2)}=B_6^{N=4,(3)}=
{15\over 4} \sum_{(H,G) \neq (0,0)}
{4+\chi [^H_G] + {\bar \chi }[^H_G]~\over 4}
{}~~\sum_{\gamma, \delta}~
|\theta[^{\gamma}_{\delta}]~|^2
{}~|\theta[^{\gamma+H}_{\delta+G}]~|^2 ~~~~~~~~~~
$$
\bea
&&B_6^{N=6,(1)}=B_6^{N=6,(2)}=B_6^{N=6,(3)}={45\over 8 }
 \sum_{(h,g)\neq (0,0)}~{{\bar \chi} [^h_g]\over
2}~\Gamma_{2,2}[^h_g]~|_{T=U=i}
\cr
&&B_6^{N=6,(4)}={45\over 8 }
 \sum_{(h,g)\neq (0,0)}~{{\chi} [^h_g]\over
2}~\Gamma_{2,2}[^h_g]~|_{T=U=i}~.
\eea

In the limit ${\rm Im}~\tau \rightarrow\infty$,
$B_6^{\rm total}=B_6^{N=4,(1)}+B_6^{N=4,(2)}+B_6^{N=4,(3)}
+B_6^{N=6,(1)}+B_6^{N=6,(2)}+B_6^{N=6,(3)}+B_6^{N=6,(4)}
\rightarrow~3\times 15 + 4\times 45/8=135/2$,
which corresponds
to the contribution
of the massless fields of the  $N=3$ supergravity
together with the
contribution of  one $N=3$ massless  vector multiplet;
$B_6({\rm massless}) =525/ 8 +1\times 15/8= 135/ 2$.

In the rank 1 model all perturbative moduli are fixed.
There is no marginal deformation in this model and all $T$
and $U$ moduli are fixed in all complex planes; in particular
the $T$ moduli are fixed to their self-dual values $T=i$.

In the next section we extend these results and write down
the non-perturbative BPS formula for $N=3$ string theory.
These results can be used for computing the one-loop
corrections to the $R^4$ and $R^6$ terms in string theory
as well as for verifying its duality with the
heterotic string theory presented in the next section.

\sxn{ Non-perturbative BPS States}

\subsection{ Mapping to (3, 0) Models}
\vskip .4cm

In this section the projections, used for
(2,1)-superstring constructions in type II models,
are mapped either to the (3,0)--heterotic or (3,0) type II
theories.
The knowledge of this mapping  defines at the  non-perturbative
level a specific $Z_2$ projection reducing the supersymmetries
from $N=4$ to $N=3$  either
in (4,0)--heterotic or in (4,0)--type II
theories. As a result, the $N=3$ non perturbative  BPS mass
formula is obtained  from the one which is valid in $N=4$ theories
via the non-perturbative $Z_2$ truncation defined by the
string--string  duality maps.

First we consider the $SU(3, 11)$ model  defined in previous section.
To derive the non-perturbative BPS mass formula it is more convenient
to use the asymmetric orbifold language. In this language the
$~b^{H}_{11}$
and $b^h_{11}$ act by  twisting (asymmetrically) some of the internal
coordinates. Namely,

 \bea
     && (i) ~b^{H}_{11}: ~~~~X^{L, R}_i  \rightarrow - X^{L, R}_i,
                                           (i=5,...,8), \cr
                                       \cr
     && (ii)~ b^h_{11}: (X^{L}_{3, 4, 5,6})  \rightarrow
                   - (X^{L}_{3, 4, 5,6})   \cr \cr
                &&~~~~~   X^{L, R}_7  \rightarrow  X^{L, R}_7 + \pi
                                 \label{n3proj}
\eea
In order to keep the world-sheet supercurrent invariant, the
orbifold projections also act on the world-sheet fermions.
These actions, in this case, are identical to the ones
specified above for the bosons. It is also evident that
$b^h_{11}$ acts freely whereas the action of $b^{H}_{11}$ has
$16$ fixed points.

In section 2, several other models with different
numbers of gauge fields have been constructed. As all of them
have lower-rank gauge sector, the corresponding BPS formula
is obtained by setting to zero some of the charges
in the above model.

The projection $b^H_{11}$ gives a $(2, 2)$ supersymmetric model in
four dimensions. This model also has a space-time interpretation
directly in six dimensions and is in fact a special case of the
type II compactification on K3. In four dimensions, it has twelve
gauge fields, and associated scalars,
from the untwisted sector. Out of these, only four
associated with the momentum and winding modes of $T^2$, specified
by the compactified dimensions $X_{3, 4}$,
are from the NS--NS sector and eight from the R--R sector. In
addition,
there are sixteen $N=4$ vector multiplets from the twisted
sectors as well. These are located at the points $X_i = (0, \pi)$
along directions $i= 5,...,8$.

The projection $b^h_{11}$ applies an asymmetric twist to
break another $1/2$
supersymmetry from the left-moving sector. Moreover, it also
applies a half-shift (on a lattice vector)
in  the $X_7$  direction. As a result,
there are no extra massless states due to this projection.
At the massless level, $b^h_{11}$
projects out half of the vectors from both the untwisted
and the twisted sectors. The numbers of  vectors in the
untwisted (NS--NS as well as R--R) sectors decrease, because of the
twist part of $b^h_{11}$, since they act as a $Z_2$ which
permutes, in pairs, the eight vectors in the R--R sector. In
the NS--NS sector two vectors are even and the other two are
odd under this $Z_2$. The shift part
of $b^h_{11}$, namely $X^7 \rightarrow X^7 + \pi$ permutes
the sixteen fixed points of the first projection ($b^H_{11}$)
in eight pairs. Its action on the vectors,
through the appropriate twist operators
associated with the fixed points, is by a similar
permutation. The shift does not have any effect
on the transformation of the vectors in the untwisted sector.
Since the model constructed by a twist $b_{11}^H$ is dual by standard
string-string duality to
the heterotic string, the projection $b^h_{11}$ can be mapped
to the heterotic side.

However, before going to the  heterotic dual, we use
the string duality which is valid among
pairs of type II string constructions, and map the
projection $b^h_{11}$ to a model possessing $(4, 0)$ supersymmetry.
The duality between type II string models with
$(4, 0)$ and $(2, 2)$ supersymmetries has been discussed
earlier\cite{seva},\cite{R2corr};
There, it was shown that such type II dual pairs
can be constructed, and they are
related through an element of the $SO(5, 5)$ {\cal U}-duality
group in six dimensions\cite{seva}. In addition, the relationship
between the two models in four dimensions also involves an
interchange between
the $S$ and $T$ moduli fields associated with the $T^2$
specified by $X_{3, 4}$.

We now obtain the mapping of the orbifold element
$b^h_{11}$ into the $(4, 0)$ side in the type II theory. Later
on, this will be extended to the heterotic string theory,
through the inclusion of the twisted sector states in
the type II models. The transformation $b^h_{11}$, up to a shift,
is represented by a
six-dimensional  $O(4, 4)$ transformation, in the
choice of metric:
\bea
     \bar{L} = \pmatrix{-I_4 & \cr
                            & I_4},
\eea
as
\bea
      \bar{\Omega} = \pmatrix{-I_2 & & \cr
                          & I_2 & \cr
                          & & I_4 }.    \label{omega4}
\eea
By complexifying the six internal coordinates as:
\be
  Z_1 = X_3 + i X_4,  Z_2 = X_5 + i X_6,  Z_3 = X_7 + i X_8,
                          \label{zed}
\ee
in the notations of \cite{seva},
$\bar{\Omega}$ can also be alternatively represented as:
\be
         \bar{\Omega} = ( \pi, 0; 0, 0),    \label{pi}
\ee
where the entries in the right-hand side of (\ref{pi})
denote the rotation in the planes represented by $Z_2$ and
$Z_3$ in the left- and the right-moving sectors,
respectively; $b^h_{11}$ also has a part that acts
as an $O(2, 2)$ transformation:
\be
  \Omega_D = \pmatrix{-I_2 & \cr
                         & I_2 },    \label{omega2}
\ee
in the planes defined by $Z_1$ in the left- and right-moving
sectors,  for a choice of the $O(2, 2)$ metric:
\be
    L_D = \pmatrix{-I_2 & \cr
                        & I_2}.   \label{eld}
\ee
The subscripts in eqs. (\ref{omega2}) and (\ref{eld})
denote the choice of a diagonal metric for $O(2, 2)$.
For later convenience, we will also use an off-diagonal
metric:
\be
   L = \pmatrix{ & I_2 \cr
                I_2 & }.     \label{el2}
\ee
These are related by a map:
\be
   L_D = \eta L \eta^T,
\ee
with
\be
    \eta = {1\over \sqrt{2}}\pmatrix{-I_2 & I_2 \cr
                                     I_2 & I_2}.
\ee

It has been shown in \cite{seva} that the mapping of an
$O(4, 4)$ element to the $(4, 0)$ side involves the
use of the triality between the $SO(4, 4)$ representations,
such that $\bar{\Omega}$ transforms to
\be
    \tilde{\bar{\Omega}} = ( \pi/2, -\pi/2; \pi/2, -\pi/2).
                           \label{pi/2}
\ee
In the matrix notation, this is written explicitly
in a block diagonal form as
\be
   \tilde{\bar{\Omega}} =
      {\rm diag}~~( i\sigma_2, -i\sigma_2 ;i\sigma_2, -i\sigma_2)
                              \label{tbaromega}
\ee
with $\sigma_2$ a Pauli matrix.

In order to map $O(2, 2)$ transformation to the
$(4, 0)$ side, we use the
metric $L$ in eq. (\ref{el2}). We also use the fact that
an $O(2, 2)$ transformation $\Omega$ can be
identified with two $SL(2)$ transformations $\Lambda_T$
and $\Lambda_U$ thanks to the equivalence:
\be
      O(2, 2) \equiv SL(2)_T \times SL(2)_U.
\ee
In particular, when the $SL(2, Z)$ transformations for the moduli $T$
and
$U$ are given as:
\bea
    \Lambda_T = \pmatrix{p_1 & q_1 \cr
             r_1 & s_1 }\in {SL(2, Z)}_T, \>\>\>\>
    \Lambda_U = \pmatrix{p_2 & q_2 \cr
             r_2 & s_2 }\in {SL(2, Z)}_U,
\eea
the $O(2, 2)$ transformation is parametrized as\cite{seva}
\bea
\Omega = \pmatrix{ p_1 p_2 & p_1 q_2 & -q_1 q_2 & q_1 p_2 \cr
                     p_1 r_2 & p_1 s_2 & -q_1 s_2 & q_1 r_2 \cr
                    -r_1 r_2 &-r_1 s_2 &  s_1 s_2 &-s_1 r_2 \cr
                     r_1 p_2 & r_1 q_2 & -s_1 q_2 & s_1 p_2 }.
                    \label{o22}
\eea
The transformation to the diagonal metric of the form described in
eq. (\ref{eld})
is through a map:
\be
\Omega_D = \eta \Omega \eta^T.
\ee
It can then be verified that the $O(2, 2)$ transformation
(\ref{omega2}), associated with the action of the
projection element in the two-dimensional space
$X_{3, 4}$ can be identified with the $SL(2)_T$ and
$SL(2)_U$ elements:
\be
   \Lambda_T = i\sigma_2, \>\>\> \Lambda_U = -i\sigma_2.
\ee
For completeness, we also mention that
the $N=3$ type II construction being completely perturbative
in nature, the $S$-duality element associated with
$b^h_{11}$ is trivial:
\be
   \Lambda_S = I_2.
\ee

Now an $S\leftrightarrow T$ interchange, together
with $U \rightarrow U$,
implies that the projection elements: $\Lambda_T$, $\Lambda_U$
and $\Lambda_S$ transform in the $(4, 0)$ side to
\be
 \tilde{\Lambda}_T = I_2, \>\>\> \tilde{\Lambda}_S = i\sigma_2,
\>\>~~~{\rm and}~~~\>\>\tilde{\Lambda}_U = -i\sigma_2 .
\label{tlambda}
\ee
As a result the projection element in the $(4, 0)$ side,
$\tilde{b^h_{11}}$, now has a non-trivial
$S$-duality action. The action of $\tilde{\Lambda}_T$ and
$\tilde{\Lambda}_U$ is further combined into an
$O(2, 2)$ projection; for the diagonal choice of the metric
$L$, this now has the form:
\be
   \tilde{\Omega}_D = \pmatrix{-i\sigma_2 & \cr
                                     & -i\sigma_2}.  \label{tomega}
\ee

The difference in the
form of the $O(4, 4)$ part of the projection elements
$b^h_{11}$ and $\tilde{b^h_{11}}$ in the $(2, 2)$ and the
$(4, 0)$ side can be understood from the fact that the
gauge-field sectors on the two sides are related through an
interchange of the NS--NS gauge fields with the R-R ones.
This is essentially a change from the vector to the spinor
representation of $SO(4, 4)$ and leads to a form of
$\tilde{b^h_{11}}$ seen above. On the other hand, the remaining
action of the projections in the two sides has its origin
in the $S\leftrightarrow T$ interchange,
which has also been used in showing the $SL(2, Z)$ duality of the
heterotic string starting from the string/string-duality
conjecture in six dimensions.

To promote the above mapping to the full heterotic string
theory, one has to analyse the action of $b^h_{11}$ on the
twisted sector states in the type II side with $(2, 2)$
supersymmetry and use a mapping of the massless fields
from the type II (on $K3$) to the heterotic string
(on $T^4$). This map transforms the various
6-dimensional fields as\cite{sensol}:
\bea
    \phi' = -\phi,\>\>\>&G'_{\mu \nu} = e^{-\phi}G_{\mu \nu}, \cr
    M' = M, \>\>\>\>\>\>&A'^{a}_{\mu} = A^{a}_{\mu}, \cr
    \sqrt{-G}e^{-\phi}H^{\mu \nu \rho}& =
    {1\over 6} \epsilon^{\mu \nu \rho \sigma \tau \epsilon}
                            H'_{\sigma \tau \epsilon} ,
                                    \label{hetmap}
\eea
where prime and unprime variables denote the fields in the
type II side, compactified on $K3$ and the heterotic side,
compactified on $T^4$.

Now, in the type II side with $(2, 2)$ supersymmetry,
    the action of $b^h_{11}$ on the sixteen vectors
from the twisted sectors of $b^H_{11}$
 can be written as an O(16) matrix in a block diagonal form:
\be
\tilde{\Omega}_T = {\rm diag} (i\sigma_2, i\sigma_2, ..., i\sigma_2).
                               \label{omegat}
\ee
This is because the
shift part of the projection $b^h_{11}$ transforms the
$X_7 = 0$ fixed points to that of $X_7 = \pi$ and
vice versa. The mapping of these
transformations to the heterotic side is then achieved
through eq. (\ref{hetmap}). The relative minus sign
in the transformation of the twisted sectors, within
a pair, can be
explained from the fact that, although the twist
fields associated with the vertex operators of these
gauge fields have identical weights at the  fixed points
$X^7 = 0$ and $X^7 = \pi$, the $U(1)$ vacuum charges are
opposite and give a relative minus sign in the
transformations.

Since the projection $b^h_{11}$ acts freely, the twisted
sector states, from this particular projection,
in the $(2, 2)$ side are heavy. This also
holds in the $(4, 0)$ side, where the projection
$\tilde{b^h_{11}}$ introduces non-zero R--R gauge field
flux at the new fixed points and makes them heavy.

We have therefore identified, through the type II/heterotic map,
the appropriate transformations in the heterotic side, which will
give an $N=3$ supersymmetric model. We notice that, unlike in the
type II side, in the heterotic case, the projection acts
non-perturbatively. However, this map does define a consistent
model and allows us to write down the expression for the
masses of the BPS states. We would once again like to
mention the similarity between this $N=3$ model and the
orbifold limit of certain $F$-theory constructions. In the
$F$-theory context, the orbifold limit is identified with
the degeneration of the fibre through appropriate Weierstrass
equations, and in this limit the moduli on the base remain
constant, modulo the monodromies around certain points, which
are identified with the fixed points of the orbifold group.
A $Z_4$ projection, which forces the 10-dimensional
type IIB axion-dilaton
moduli to be fixed to its value at the self-dual point,
was also identified in the $F$-theory context\cite{dasmuk}.
It may
be of interest to examine whether such a similarity
between the two cases also leads to an understanding of
certain non-perturbative aspects in our case.

We conclude this subsection by summarizing the action of
the $N=3$ projection $\tilde{b^h_{11}}$ in the heterotic side.
This is given by eqs. (\ref{tlambda}), (\ref{tomega}) on the
coordinates $X_{3,4}$, by $\tilde{\bar{\Omega}}$
in eq. (\ref{tbaromega}) on the coordinates $X_{5,6,7,8}$, and
by $\tilde{\Omega}_T$ in eq. (\ref{omegat}) on the extra sixteen
right-moving coordinates of the heterotic string.
Since these projections act as an exchange of
internal coordinates,
they reduce the rank of the gauge group by half. In the
next subsection we will see the same phenomena from
a different point of view. The $S$-duality projection
$\tilde{\Lambda}_S$ will be shown to preserve only
 half of an $N=4$ vector
multiplet. These exchange operations then
combine two such half-multiplets to construct a
self-conjugate vector multiplet of $N=3$.

\subsection{Massless States}
\vskip .4cm

After having identified the projection in the heterotic side, we now
obtain the spectrum of the heterotic string theory, when the
above projection is applied.
We show that there are three surviving supersymmetries, and
at the massless level, we get a correct
spectrum for the $N=3$ string theory. In an
$N=4$ theory, the supersymmetry algebra and
representations are specified by a group structure,
$SO(2)\times SU(4)\times U(1)_S$, where $SO(2)$ is the little
group of the Lorentz group and $U(4) \equiv SU(4) \times U(1)_S$
is the $R$-symmetry for the $N=4$ theory. Gravitinos transform
as ${\bf 4}$ and the $U(1)$ gauge fields of the
supergravity multiplet transform as a ${\bf 6}$ of $SU(4)$.
In addition, $N=4$ supergravity multiplet also has two scalars,
which are neutral under $SU(4)$. In heterotic string theory,
the $SU(4)$ symmetry
can be identified with the $SO(6)_L$ subgroup of the
$SO(6, 22)$ $T$-duality group, which originates from its
left-moving sector.
$U(1)_S$ is the maximal compact subgroup of the
$S$-duality group: $SL(2, R)$.

For the projection $\tilde{b^h_{11}}$, the $SU(4)$ element is given
by
the diagonal $U(1)$ subgroup of $SO(6)_L$:
$\Omega_L \equiv {\rm diag} (-i\sigma_2, i\sigma_2, -i\sigma_2)$.
In this case, we also have, in the right-moving sector, a
projection given by the diagonal $U(1)$ element of $SO(22)_R$,
given by $\Omega_T$ in eq.~(\ref{omegat}) together with
$\Omega_R$, which acts on the
right-moving part in the 6-dimensional internal space and is
identical to $\Omega_L$. In addition, as mentioned before
the $U(1)_S$ element is specified by the
matrix $\tilde{\Lambda}_S$ given in eq. (\ref{tlambda}).

This projection breaks $SU(4)\times U(1)_S$ to its subgroup
$SU(3)\times U(1)_V$, where the $U(1)_V$ that remains unbroken is
a combination of the $U(1)_S$ and $U(1)_I$, with $U(1)_I$
as the diagonal $U(1)$ subgroup of $SU(4)$:
$SU(3) \times U(1)_I \subset SU(4)$.
The four supercharges and their CPT conjugates transform,
under $SU(4)\times U(1)_S$ as:
\be
   Q_{1/2} =  {\bf 4}_{(1/2, 1)} + \bar{\bf 4}_{(-{1/2}, -1)},
                                  \label{supercharge}
\ee
where the first entry in the bracket shows the helicity
of the state.
By decomposing the supercharges in the representation
of $SU(3)\times U(1)_I$, the three supersymmetry generators that
survive the projection are: $ {\bf 3}_{[{1/2}, -{1/3}, 1]}$ and
its complex conjugate.
 In the following, the entries in the curly
brackets ``$(\cdot,\cdot)$" denote helicity and $U(1)_S$ charges
while the
quantities in the square brackets ``$[\cdot,\cdot,\cdot]$" denote
helicity, $U(1)_I$
and $U(1)_S$ charges.
 The supercharge transforming as a
singlet of $SU(3)$ is projected out. As a result we are
left with $N=3$ supersymmetry.

The states that survive the above projection also belong to a
representation of the residual supersymmetry. The $N=4$ gravity
multiplet is constituted out of helicity $\pm 2$ supermultiplets of
the $N=4$ supersymmetry, which are complex conjugates
of each other. They have
the transformation property:
${\bf 1}_{(-2, 0)} + {\bf 4}_{(-{3/2}, 1)}
+ {\bf 6}_{(-1, 2)} + \bar{\bf 4}_{(-{1/2}, 3)}
+ {\bf 1}_{(0, 4)}$, together
with the complex-conjugate representation. The $N=3$ projection
then selects out of the above,
the ${\bf 1}_{[2, 0, 0]}$ and
$\bar{\bf 3}_{[-1, -{2/3}, 2]}$ states among the
bosonic ones, due to the decomposition of
$SU(4)\rightarrow SU(3) \times U(1)_I $.  These, together with their
CPT conjugates, give us
$g_{\mu \nu}$ and three $A_{\mu}$'s, which is the correct spectrum
for the $N=3$ supergravity sector.

The vector multiplet of $N=4$ supersymmetry is self-conjugate
under CPT, whereas the $N=3$ one is constructed out of two
different multiplets  containing helicity $-1$ and $+1$ states
and are conjugate to each other. The $N=3$ projection mentioned
above does not leave the complete vector multiplet of the
$N=4$ theory invariant. Instead, it projects out some of the
states and leaves only an $N=3$ multiplet, of either  $+1$
or $-1$ helicity invariance. However, as seen before,
the $N=3$ projection   also permutes the
internal indices of the heterotic string theory in the
right-moving sector. As a result, the complete $N=3$ vector multiplet
is a linear combination of two half-vectors from $N=4$. The
final spectrum is CPT-invariant. But the rank of the gauge group is
reduced by a factor $1/2$.

\subsection{Mass Formula}
\vskip .4 cm

After identifying the correct $N=3$ massless spectra from the
projection of the heterotic strings, we now proceed to
write down the BPS mass formula in this theory. In this context,
the starting point is the $N=4$ BPS formula, which can be
written in terms of six ``electric'' and six ``magnetic'' charges,
associated with the supergravity sector, as\cite{kiritsis}:
\be
    M^2_{BPS}   = { {[(P_m + S Q_m)
                  (P^m + \bar{S} Q^m)]}\over 4 ~{\rm Im}~ S}
                 + {1\over 2} \sqrt{(P_m P^m) (Q_m Q^m)
                   - {(P_m Q^m)}^2},
                                  \label{n4bps}
\ee
where the contractions of the indices are defined with
respect to the internal metric on $T^6$, namely $G^{m n}$.
The quantities $P_m$ and $Q_m$ are defined in terms of the
integer valued electric charges $(\alpha^L, \alpha^R, \alpha^I)$
and magnetic charges
$(\tilde{\alpha}^L, \tilde{\alpha}^R, \tilde{\alpha}^I)$
of the heterotic string theory as:
\bea
 P & = \alpha^L + (G + B + C) \alpha^R + A \alpha^I  \cr
 Q & = \tilde{\alpha}^L +
        (G + B + C) \tilde{\alpha}^R + A \tilde{\alpha}^I.
                              \label{charge}
\eea
Here $C = {1\over 2} A A^T$,
$G$ and $B$ are the moduli fields associated with the
internal $T^6$, and the $A$'s are the Wilson-line moduli.

The square-root factor in the BPS formula is proportional to the
square of the difference in the two $N=4$ central charges.
This term vanishes for  the
states preserving $1/2$ supersymmetry. Such states belong to the
``short'' multiplets of the $N=4$. The  non-zero
contribution comes from the states belonging to
the intermediate multiplets of $N=4$ and  preserve the $1/4$ of the
supersymmetry.
The BPS states in the perturbative construction of the
heterotic string theory are  examples of ``short'' multiplets.

To obtain the BPS formula for the
$N=3$ case, using the $Z_2$ projection mentioned before,
we must set at the beginning first
the ``dilaton-axion'' moduli of the $N=4$ theory
to the value $S=i$.
This is due to an observation that we made earlier,
namely a transformation by $\tilde{\Lambda}_S$ in
eq. (\ref{tlambda}) is a symmetry of the $N=4$ theory,
for a given coupling, only for this value of the $S$ field.
In other words, the $N=3$
projection transforms the strong coupling to the weak one, and
can therefore be applied consistently
only for its fixed value at the self-dual point.
Furthermore, since the gauge fields,
associated with the charges mentioned above, transform under
$SU(4)\times U(1)_S$ as ${\bf 6}_{(-1, 2)}$ and ${\bf 6}_{(1, -2)}$,
only those belonging to the $SU(3)\times U(1)_V$ representations,
${\bf 3}_{[1, 2/3, -2]}$ and
$\bar{\bf 3}_{[-1, -{2/3}, 2]}$, survive the $N=3$ projection. To
select
the appropriate combinations that remain invariant under this
projection, we define:
\be
      \Pi_m = (P_m + i Q_m), \>\>\>\>
      \bar{\Pi}_{\bar m} = (P_m - i Q_m).      \label{pim}
\ee
We also use the complexifications of the coordinates
introduced in eq. (\ref{zed}) and note that out of the
original twelve charges, those
existing after the projection are six charges $\Pi_{m}$
and $\bar{\Pi}_{\bar m }$.  With this projection, both the terms
in the $N=4$ mass formula give identical contributions, and one gets
\be
     M^2_{BPS} = {1\over 2}~\Pi_{m}~ G^{m \bar{m}}~ \bar{\Pi}_{\bar
m}.
                                \label{bps3}
\ee
In obtaining the $N=3$ mass formula from $N=4$ in
eq. (\ref{bps3}), we have also used the fact that the
internal space for $N=3$ is parametrized by a K\"ahler metric.
This can be observed independently of the action of
$\tilde{b^h_{11}}$ on the moduli fields. In particular, only the
components $G_{m \bar{m}}$, $B_{m \bar{m}}$,
$A_{m}^{I^+}$ and $A_{\bar{m}}^{I^-}$, with $I^{\pm}$
being the complexifications of the sixteen right-moving
coordinates, survive the $N=3$ projection. The mass formula
(\ref{bps3}) is also the unique quadratic invariant of the
charges, the latter transforming as ${\bf 3}$ and $\bar{\bf 3}$ under
the
residual $U(3)$ symmetry.

We now rewrite the BPS mass formula (\ref{bps3}) in terms
of the physical charges associated
with the gauge fields in the theory. This is also given by
a projection over the charges $\Pi_m$, defined in terms of
the physical charges of $N=4$ theory, using eqs.~(\ref{charge})
and (\ref{pim}). The final expression has a form similar to
that of eq. (\ref{charge}):
\be
    \Pi_{m}  = {\alpha^L}_{m} +
             (G + B + C )_{m \bar{m}} \alpha^R_{\bar{m}}
             + A_{m}^{I^+} \alpha_{I^+}~.
                                   \label{pii}
\ee
The mass formula (\ref{bps3})
can now be rewritten in an $SU(3, n)$-invariant form{\footnote{
In the present case $n = 11$. But the general structure of the
invariant expressions is preserved for other values of $n$
as well.}}:
\be
   M^2_{BPS} = q^{\dagger} \cdot (M + L ) \cdot q,
\ee
with $q$ denoting the column vector:
\be
      q = \pmatrix{ \alpha^L_{m} \cr
                    \alpha^R_{m} \cr
                    \alpha^{I^+}}.       \label{que}
\ee
The $SU(3, n)$ matrix $M$ has the standard expression
in terms of the Wilson-line moduli $A_{z_i}^{I^{\pm}}$ and
the Hermitian and anti-Hermitian matrices $G$ and $B$,
respectively:
\bea
    \pmatrix{G^{-1} & G^{-1} (B+C) & G^{-1} A \cr
             (-B + C)G^{-1} & (G - B + C)
             G^{-1} (G + B + C) &
             (G-B + C )G^{-1} A \cr
             A^{\dagger} G^{-1} &
             A^{\dagger} G^{-1} (G + B + C) &
             {\bf I_8} + A^{\dagger} G^{-1} A},
                                    \label{defm}
\eea
with $C={1\over 2}AA^T$   and the $SU(3, n)$ metric $L$
having the form:
\bea
     L = \pmatrix{ 0 & I_3 & 0 \cr
                  I_3 & 0 & 0 \cr
                  0 & 0 & -I_8}.      \label{el3}
\eea
It can be verified that the matrix $M$ is Hermitian and
satisfies the $SU(3, n)$ property:
$M^{\dagger} L M  = L$.

In this section we have presented the $SU(3, n)$-invariant BPS
mass formula for the $N=3$ string theories in four dimensions.
There can be many applications of these results,
including   black-hole physics. The entropy formula
for the $N=3$ case has  already been presented in the
literature\cite{ferrara}.
Our results can be used to obtain these expressions from a
microscopic description through an appropriate truncation of
the type II or heterotic string models.

\sxn{$N=3$ String Effective Action}

The projections applied to the $N=4$ theory, in the previous
section, can also be used for writing down the
$N=3$ effective action. By restricting this to the bosonic
sector, the $N=4$ effective action,
at a generic point $\hat{M}$ in the moduli space of the heterotic
string has a form:
\bea
   \hat{S} & = {1\over 32 \pi} \int \sqrt{-g}\left[ R -
     {1\over {2 (\lambda)^2}}
     g^{\mu \nu} \pa_{\mu} \lambda \pa_{\nu} \bar{\lambda}
       -  \lambda_2 F^{(a)}_{\mu \nu}
      (\hat{L} \hat{M} \hat{L}) F^{(b) ~ \mu \nu} +     \right. \cr
     & \lambda_1 F^{(a)}_{\mu \nu} \hat{L}_{a b} {\tilde F}^{(b)~ \mu
\nu} +
     {1\over 8} g^{\mu \nu}
\left.     {\rm Tr}(\pa_{\mu}\hat{M} \hat{L} \pa_{\nu}\hat{M}
\hat{L})
\right],
                                               \label{eff4}
\eea
where $(a = 1,...,28)$ and $\hat{L}$ is an $SO(6, 22)$ metric:
\be
   \hat{L} = \pmatrix{0 & I_6 & 0 \cr
                      I_6 & 0  & 0\cr
                      0 & 0 & -I_{16}}.       \label{hatl}
\ee
By taking into account the fact that the axion-dilaton moduli of the
heterotic string are fixed to the self-dual point, the
$N=3$ projection then leads to an action of the form:
\be
   S= {1\over 32 \pi} \int d^4 x \sqrt{-g} \left[ R -
     F^+_{\mu \nu} (L M L) F^{-  \mu \nu} +
     {1\over 8} g^{\mu \nu} {\rm Tr}(\pa_{\mu}M L \pa_{\nu}M L)
\right]~,
                                \label{action3}
\ee
where $M$ denotes the $SU(3, n)$ moduli (\ref{defm}), $L$ is the
$SU(3, n)$ metric (\ref{el3}) and $F^{\pm} = F \pm i \tilde{F}$.
In obtaining action (\ref{action3}) from
(\ref{eff4}), we expanded various terms of the $N=4$ action
and then recombine them after collecting the invariants.

It is observed that the action (\ref{action3})
is manifestly invariant under
the $U(3, n)$ symmetry. The manifest invariance of the action is
due to the fact that the $N=3$ projection on the heterotic
strings leaves only perturbative moduli in the spectrum.
The above reduction of $N=4$ effective action to $N=3$
can also be seen at the level of the
equations of motion, which for the $N=3$ case can be
written, starting from the $N=4$ one, as:
\bea
     R_{\mu \nu} &=& 2 F^+_{\mu \nu} ( L M L ) F_{\nu}^{\rho}
                   - {1\over 2} g_{\mu \nu}
                   F_{\rho \sigma}^+ ( LML) F^{- \rho \sigma} \cr
      D_{\mu}( MLF^{+ \mu \nu}) &=& 0
                                        \label{eom}
\eea

It will be of interest to study the solutions of these
equations of motion, in order to obtain the classical
background configurations that are consistent with
$N=3$ supersymmetry. Among them,
 those that preserve $1/2$ of the supersymmetry,
such as extremal black-hole configurations, are of particular
interest.
They will
provide the examples of the $N=3$ BPS states found in
the previous sections.

\sxn{Conclusions}

We have presented an explicit expression for the $N=3$
BPS formula. It was also shown how the known $N=3$ models
are incorporated in this picture. The BPS states associated with
the perturbative spectrum of the known $N=3$ string theory were
also described. It will be of interest to  further study
compactification
of the $N=3$ effective action to 2- and 3-dimensional
space-times. In particular, it is expected that the effective
action in three dimensions will possess an $SU(4, n+1)$
symmetry. The coset structure
$SU(4, n+1)/SU(4)\times SU(n+1) \times U(1)$ can be seen by a
direct counting of the matter degrees of freedom, which in the
bosonic sector contains only scalars, in three dimensions.
Our results, through a duality between the type II and the
heterotic sides,
also indicate that type II string models
with $N=3$ supersymmetry and non-Abelian gauge symmetries
can be obtained at special points of the moduli space.
It may be interesting to examine whether some of these
symmetry enhancements in the type II case can take
place at special values of the perturbative type II
moduli as well. One will then be able to study them
using conformal field-theory techniques.

It will also be of interest to directly construct $N=3$
orientifold models in four dimensions, whose open string
sectors, with Dirichlet boundary conditions, can be
interpreted as BPS states appearing in the mass formulae
derived earlier. This turns out to be a
difficult exercise due to the asymmetric nature of
the $N=3$ construction on the one hand and to
the requirement of the left-right symmetry for the
orientifolding operation on the other. In order to achieve the
desired results, the
orientifolding operation must therefore be combined with
appropriate operation on the
internal space.
Although the orientifolding operation seems difficult in four
dimensional models, it becomes much simpler in two dimensions
where models preserving
$3/8$ supersymmetry
can be easily constructed.

\vspace*{.4cm}

\noindent
{\bf Acknowledgements}
\vspace*{.3cm}

We would like to thank I. Antoniadis,
S. Ferrara, E. Kiritsis, B. Pioline and A. Sen for valuable discussions.
A.K. thanks the Theory Division at CERN for
its warm hospitality.
The work of C.K. is supported by the TMR contract
ERB-4061-PL-95-0789.
\vfil
\eject

\end{document}